# Astro2020 Science White Paper

# Radio sources in next-generation CMB surveys

**Thematic Areas**:

☒ Cosmology and Fundamental Physics

☒ Galaxy Evolution

**Principal Author:**
Name: Gianfranco De Zotti
Institution: INAF-Osservatorio Astronomico di Padova
Email: gianfranco.dezotti@inaf.it
Phone: +390498293444

**Co-authors:** M. Bonato (INAF-IRA, Bologna, Italy), M. Negrello (School of Physics and Astronomy, Cardiff University, UK), D. Herranz (Instituto de Física, Universidad de Cantabria, Santander, Spain), M. López-Caniego (ESAC, Villafranca del Castillo, Spain), T. Trombetti, C. Burigana and M. Massardi (INAF-IRA, Bologna, Italy), L. Bonavera and J. González-Nuevo (Departamento de Fìsica, Universidad de Oviedo, Spain), V. Galluzzi (INAF-OATs, Trieste, Italy), S. Hanany (University of Minnesota)

**Abstract**: CMB surveys provide, *for free*, blindly selected samples of extragalactic radio sources at much higher frequencies than traditional radio surveys. Next-generation, ground-based CMB experiments with arcmin resolution at mm wavelengths will provide samples of thousands radio sources allowing the investigation of the evolutionary properties of blazar populations, the study of the earliest and latest stages of radio activity, the discovery of rare phenomena and of new transient sources and events. Space-borne experiments will extend to sub-mm wavelengths the determinations of the SEDs of many hundreds of blazars, in temperature and in polarization, allowing us to investigate the flow and the structure of relativistic jets close to their base, and the electron acceleration mechanisms. A real breakthrough will be achieved in the caracterization of the polarization properties. The first direct counts in polarization will be obtained, enabling a solid assessment of the extra-galactic source contamination of CMB maps and allowing us to understand struc-ture and intensity of magnetic fields, particle densities and structures of emitting regions close to the base of the jet.

**Blazar physics**

Although a substantial progress on the characterization of mm and sub-mmm properties of extragalactic radio sources has been made in recent years mainly thanks to surveys with WMAP, *Planck*, the SPT and the ACT, the available information is still limited. The overwhelming majority of extragalactic radio sources detectable in the frequency range of CMB experiments are blazars, i.e. sources whose radio emission is dominated by relativistic jets collimated by intense magnetic fields and closely aligned with the line of sight. These objects with extreme properties are of special interest since they are also strong γ–ray sources: about 90% of the firmly identified extragalactic γ–ray sources are blazars.

Accurate source counts over large flux density intervals provide key constraints on evolutionary models of these sources. Just because high frequency surveys are still far less extensive than those at low frequencies, evolutionary models for blazar populations, Flat Spectrum Radio Quasars (FSRQs) and BL Lacertae sources (BL Lacs), are far less advanced than those for steep-spectrum radio sources. For example, while clear evidence for ``downsizing'' was reported in the case of steep-spectrum sources (Massardi et al. 2010, Rigby et al. 2015), the available data are insufficient to test if this is the case also for FSRQs; for BL Lacs the constraints on evolutionary parameters are even weaker. This situation hampers sharp tests of unified models.

*Planck* has already provided strong indications of the crucial role of blazar photometry up to sub-mm wavelengths to get information on the energy spectrum of relativistic electrons responsible for the synchrotron emission, with interesting implications for the acceleration mechanisms (Planck Collaboration XLV, 2016).

Another interesting open question is the geometry of the emitting regions. The most commonly used model for the spectral energy distribution (SED) of compact, radio loud Active Galactic Nuclei (AGNs) is a leptonic, one-zone model, where the emission originates in a single component. The SEDs typically consist of two broad-band bumps; the one at lower frequencies is attributed to synchrotron radiation while the second, peaking at gamma-ray energies, is attributed to inverse Compton. The one-zone model is generally found to provide an adequate approximation primarily because of the limited observational characterization of the synchrotron SED, with fragmentary data over a limited frequency range. However VLBI images show multiple knots often called ``components'' of the jet. The standard model (Marscher & Gear 1985) interprets the knots as due to shocks that enhance the local synchrotron emission.

The spectrum is explained as the result of the superposition of different synchrotron self-absorbed components in a conical geometry. The synchrotron self-absorption optical depth scales as $\tau_{sync} \propto B_\perp^{(p+2)/2} \nu^{-(p+4)/2}$ where $B_\perp$ is the magnetic field component perpendicular to the electron velocity and p is spectral index of the energy distribution of relativistic electrons (typically, p ~ 2.5). Thus $\tau_{sync}$ increases towards the nucleus as the magnetic field intensity and its ordering increases, but is strongly frequency dependent: the emission at higher and higher frequencies comes from smaller and smaller distances from the central engine.

Thus the mm and sub-mm emissions provide information on the innermost regions of the jets, where it is optically thin, while the emission at longer wavelengths is affected by self-absorption. Interestingly, León-Tavares et al. (2011) and Fuhrmann et al. (2016) found a significant correlation between simultaneous gamma-ray fluxes and millimeter-wave flux densities of FSRQs. The strongest gamma-ray flares were found to occur during the rising/peaking stages of millimeter flares. This suggests that the gamma-ray flares originate in the millimeter-wave emitting regions of these sources.

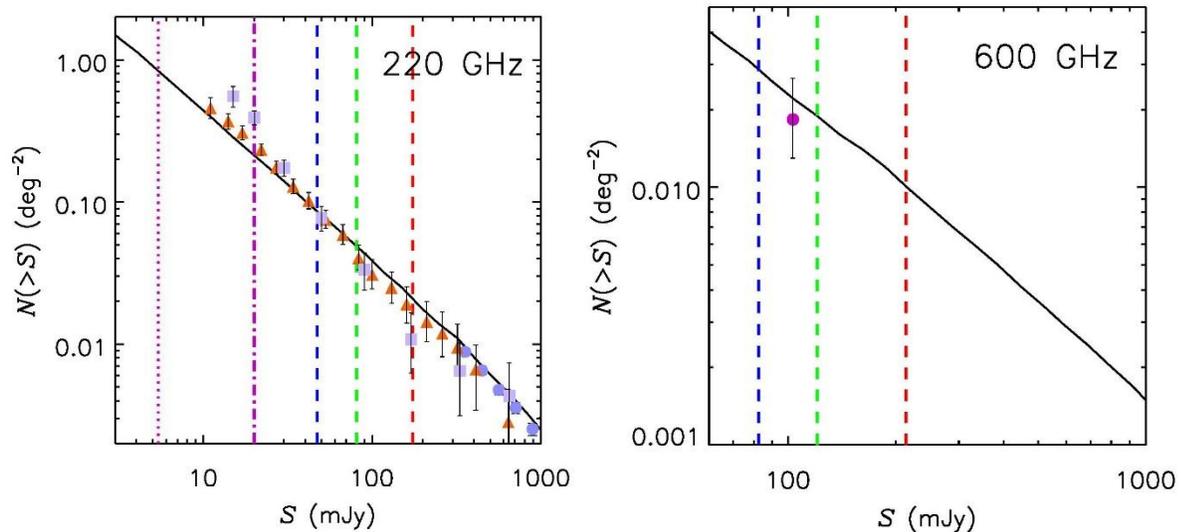

Figure 1. Integral number counts of radio sources at 220 and 600 GHz. The vertical dashed lines show the $5\sigma$ confusion limits for a space-borne experiment with a 1 m, 1.5 m and a 2 m telescope (from right to left) operating at the diffraction limit. The dot-dashed and dotted vertical lines on the left panel show the completeness limit of the SPT survey and the $5\sigma$ confusion limit for the SPT telescope, respectively. The data points on the left panel are from Mocanu et al. (2013; SPT, orange triangles), Marsden et al. (2014; ACT, lavender squares), Planck Collaboration XIII (2011; light blue circles); that on the right panel is from Bonato et al. (2019) and is based on Herschel survey data. The solid black lines are predictions of the Tucci et al. (2011) model.

The available data are mostly at cm (or longer) wavelengths and are scanty at (sub-)mm wavelengths because of the sky limited areas covered by the available surveys. Next-generation space-borne CMB experiments with ≈ 1.5 m telescopes, like PICO (Hanany et al. 2019), will fill this gap (Fig. 1). Ground-based experiments with ≈ 6 m telescopes, like CMB-S4 (Abazajian et al. 2016) and the Simons Observatory (Ade et al. 2019), will detect thousands of blazars per sr at millimeter wavelengths. Space-borne experiments will extend the spectral coverage, providing, for the first time, samples of hundreds of blazars blindly selected at sub-mm wavelengths. An important property of surveys from space is that they provide simultaneous photometry over a broad frequency range, thus overcoming the complications due to variability and allowing us to directly connect the observed SED to the physical processes operating along the jet.

**Earliest and latest phases of radio activity**
Large-area surveys at frequencies of tens to hundreds of GHz will also detect the rare but very interesting sources associated to the earliest and to the latest stages of the radio-AGN evolution, both characterized by emissions peaking in this frequency range (De Zotti et al. 2005).

It is now widely agreed that extreme GHz peaked spectrum (GPS) sources correspond to the early stages of the evolution of powerful radio sources, when the radio emitting region grows and expands within the interstellar medium of the host galaxy, before plunging in the inter-galactic medium and becoming an extended radio source. There is a clear anti-correlation between the peak (turnover) frequency and the projected linear size of GPS sources, suggesting a decrease of the peak frequency as the emitting blob expands. The identification of these sources is therefore a key element in the study of the early evolution of radio AGNs. High-frequency surveys will detect these sources very close to the moment when they turn on.

Possible examples of extremely young sources are the six narrow-line Seyfert 1 galaxies detected by Lähteenmäki et al. (2018) at 37 GHz with flux densities of 270-970 mJy but undetected by the FIRST survey, complete down to ~1 mJy at 1.4 GHz, carried out about 20 yr ago. One possibility is that the new observations have discovered newly triggered radio activity from nuclei that were essentially radio silent two decades ago. If so, the multifrequency surveys by next generation CMB experiments will enable studies of the launching of relativistic jets, as well as of the evolutionary paths that young AGNs take on their way to becoming fully-evolved, powerful radio sources. Other possibilities have been considered. In any case, these are very interesting objects. The next generation CMB surveys will allow us to determine their abundance and will measure their SEDs, shedding light on their nature.

Large area (sub-)mm surveys will also allow us to investigate the late stages of the AGN evolution in elliptical galaxies, characterized by low radiation/accretion efficiency. These manifest themselves via a nuclear radio emission described by advection-dominated accretion flows (ADAFs) and/or by adiabatic inflow-outflow solutions (ADIOS). Doi et al. (2005) have found that at least half of their sample of 20 low-luminosity active galactic nuclei with compact radio cores show radio spectra rising at least up to 96 GHz, consistent with the `sub-millimetre bump' predicted by an ADAF model. Again CMB surveys will determine the abundance of these objects and their SED measurements will clarify the origin of the emission.

The spectral range covered by CMB experiments will also allow the study of the connection between radio activity and star formation. *Planck* has detected evidences of cold dust associated to a handful of nearby radio sources, based on their rising spectra at mm wavelengths. The much deeper surveys carried out by next-generation experiments can push the investigation to much more distant objects.

Predictions of the expected number of detections of these source populations are limited to $\leq$ 30 GHz (De Zotti et al. 2005, Tinti & De Zotti 2006). They suggest that at 20-30 GHz hundreds of these objects can be detected by ground-based experiments. Space-borne experiments are confusion-limited to much brighter flux densities, implying a detection rate at least one order of magnitude lower. The number of detections is predicted to drop rapidly with increasing frequency (Blandford & McKee 1976, Granot & Sari 2002).

**The transient sky**
High-sensitivity and high-angular-resolution CMB surveys also offer a unique opportunity to carry out an unbiased investigation of the largely unexplored mm/sub-mm transient sky

(Metzger et al. 2015). So far measurements have been limited to follow-up of objects detected at other wavelengths, with limited success partly because of the need for excellent weather conditions or because they were too short-lived to detect without very rapid reaction times. CMB surveys will allow us to discover new, unknown transient sources in this band.

One example of transient phenomena are outbursts from AGNs and especially from blazars. Outbursts and, more generally, variability, provide key information on the flow of the plasma within the relativistic jets. Signatures of evolving shocks in the strongest radio flares were seen by Planck Collaboration Int. XLV (2016), although much of the high frequency variability may be better approximated by achromatic variations. These results are compatible with the standard shocked jet model, but other interpretations are possible. Definite conclusions on all the above issues are currently hampered by the limited statistics. This limit will be overcome by next generation CMB experiments from space which will provide multi-epoch simultaneous observations of large blazar samples over a broad frequency range. This will allow us to study their variability properties as a function of flux density and spectral shape.

Perhaps even more interesting is the possibility of detecting radio aftergrows of gamma-ray bursts (GRBs). Afterglows often have a spectral peak in or near the mm range (Granot & Sari 2002), with emission lasting over days to weeks. One candidate object with properties broadly consistent with a GRB afterglow was tentatively detected by Whitehorn et al. (2016) on SPT data over 100 deg$^2$ with an observing time of 6,000 h, but the statistical significance of the detection was too low to completely rule out a fluctuation. GRB emission is expected to be less tightly beamed at these wavelengths than in gamma-rays. Thus afterglows not accompanied by detectable gamma-ray emission are expected to exist, but have not been detected yet. Blind surveys of large sky areas by next generation ground-based CMB experiments down to ~10 mJy sensitivity can reveal these orphan afterglows and new, unknown sources. Multiple detections per year are expected. Even a non-detection will place interesting constraints on the shock dynamics and on the energy budget of the unknown GRB progenitors.

One example of unexpected phenomena that may show up at (sub-)mm wavelengths is the extraordinary extragalactic transient AT2018cow, with an estimated peak flux density of 94 mJy at $\simeq$ 100 GHz (Ho et al. 2019). This object heralds a new class of energetic transients which at early times are most readily observed at (sub-)mm wavelengths.

**Polarimetry**

Polarimetric observations will take full advantage of the improvement in sensitivity by almost two orders of magnitude of next generation CMB experiments, compared to *Planck*. This is because at this resolution the graininess of the sky is much lower in polarization than in temperature so that the detection limit is mostly dictated by sensitivity rather than by confusion noise, although confusion becomes increasingly important with decreasing angular resolution . This was demonstrated by simulations reported by De Zotti et al. (2018) for space-borne telescopes of 1 to 1.5 m. Hence these experiments will allow a real breakthrough in the characterization of the polarization properties of extragalactic sources. So far polarimetric observations of radio sources at (sub-)mm wavelengths are limited to few samples, typically selected at cm-wavelengths.

The most extensive mm/sub-mm polarization follow-up surveys were carried out by Trippe et al. (2010) with the IRAM Plateau de Bure Interferometer and by Agudo et al. (2010, 2014) with the IRAM 30 m telescope. Most recently, Galluzzi et al. (2018) obtained, with ALMA, polarimetric observations at 97.5 GHz of a complete sample of 32 extragalactic radio sources.

The number of polarization detections in blind surveys is very limited. *Planck* detections in the ``extragalactic zone'' (approximately $|b|>20°$) range from 28 at 30 GHz to ~10 in the channels up to 217 GHz, to 1 at 353 GHz (De Zotti et al. 2018}; the two highest Planck frequencies (545 and 857 GHz) were not polarization sensitive. All detected sources are radio-loud AGNs. Estimates of the mean polarization fraction of extragalactic sources were obtained applying stacking techniques to *Planck* 30-353 GHz polarization maps (Bonavera et al. 2017a,b; Trombetti et al. 2018).

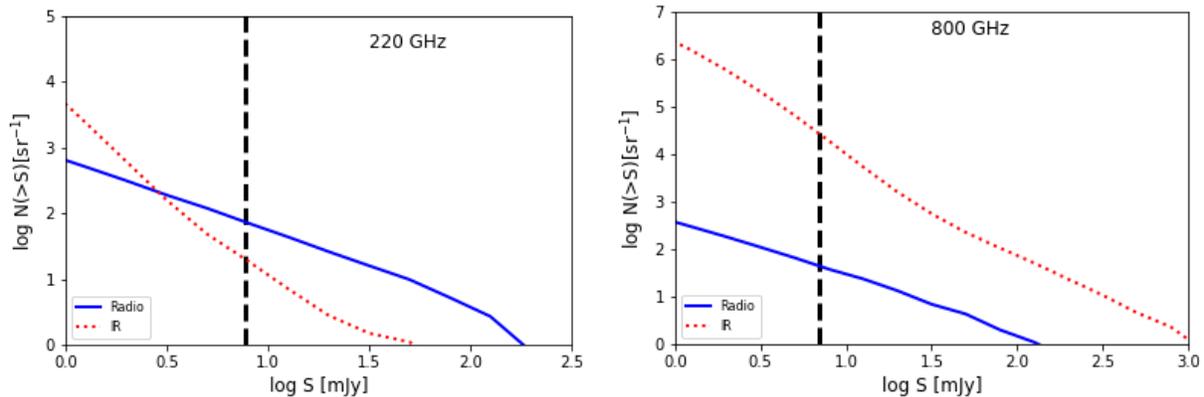

*Figure 2. Estimated integral number counts in polarized intensity of radio sources and of dusty galaxies (IR) at 220 and 800 GHz. The vertical dashed black lines show the $5\sigma$ detection limits for a space-borne instrument with a 1.5 m telescope and state-of-the-art sensitivity derived from the simulations described in De Zotti et al. (2018). These simulations assumed log-normal distributions of the polarization fractions with mean and dispersion of 2.14% and of 0.9%, respectively, for radio sources and of 1.4% and 1%, respectively, for dusty galaxies. The assumed mean polarization fraction of dusty galaxies was derived from Planck polarization maps of the Milky Way.*

As illustrated by Fig. 2, next-generation CMB experiments will provide the first direct counts of extragalactic sources in polarization at mm and sub-mm wavelengths. On one side this will allow a much better control of the extragalactic source contamination of CMB maps. This is particularly important in the 60–120 GHz frequency range, where diffuse polarized foreground emissions display a broad minimum. Accurate simulations (Remazeilles et al. 2018) showed that, for tensor-to-scalar ratios r ~$10^{-3}$, the overall uncertainty on r is dominated by foreground residuals and that unresolved polarized point sources can be the dominant foreground contaminant over a broad range of angular scales. On the other side, polarization observations enable us to understand geometrical structure and intensity of magnetic fields, particle densities and structures of emission regions.

**References**

Abazajian, K. N., Adshead, P., Ahmed, Z., et al. 2016, arXiv:1610.02743
Ade, P., Aguirre, J., Ahmed, Z., et al. 2019, JCAP, 2, 056
Agudo, I., Thum, C., Gómez, J. L., & Wiesemeyer, H. 2014, A&A, 566, A59
Agudo, I., Thum, C., Wiesemeyer, H., & Krichbaum, T. P. 2010, ApJS, 189, 1
Blandford, R. D., & McKee, C. F. 1976, Physics of Fluids, 19, 1130
Bonato, M., Liuzzo, E., Herranz, D., et al. 2019, MNRAS, 485, 1188
Bonavera, L., González-Nuevo, J., Argüeso, F., & Toffolatti, L. 2017a, MNRAS, 469, 2401
Bonavera, L., González-Nuevo, J., De Marco, B., et al. 2017b, MNRAS, 472, 628
De Zotti, G., Castex, G., González-Nuevo, J., et al. 2015, JCAP, 6, 018
De Zotti, G., González-Nuevo, J., Lopez-Caniego, M., et al. 2018, JCAP, 4, 020
De Zotti, G., Ricci, R., Mesa, D., et al. 2005, A&A, 431, 893
Doi, A., Kameno, S., Kohno, K., Nakanishi, K., & Inoue, M. 2005, MNRAS, 363, 692
Fuhrmann, L., Angelakis, E., Zensus, J. A., et al. 2016, A&A, 596, A45
Galluzzi, V., Puglisi, G., Burkutean, S., et al. 2018, MNRAS, submitted
Granot, J., & Sari, R. 2002, ApJ, 568, 820
Hanany, S., Alvarez, M., Artis, E., et al. 2019, arXiv:1902.10541
Ho, A. Y. Q., Phinney, E. S., Ravi, V., et al. 2019, ApJ, 871, 73
Lähteenmäki, A., Järvelä, E., Ramakrishnan, V., et al. 2018, A&A, 614, L1
León-Tavares, J., Valtaoja, E., Tornikoski, M., et al. 2011, A&A, 532, A146
Marscher, A. P., & Gear, W. K. 1985, ApJ, 298, 114
Marsden, D., Gralla, M., Marriage, T.~A., et al. 2014, MNRAS, 439, 1556
Massardi, M., Bonaldi, A., Negrello, M., et al. 2010, MNRAS, 404, 532
Metzger, B. D., Williams, P. K. G., & Berger, E. 2015, ApJ, 806, 224
Mocanu, L. M., Crawford, T. M., Vieira, J. D., et al. 2013, ApJ, 779, 61
Planck Collaboration XIII 2011, A&A, 536, A13
Planck Collaboration XLV 2016, A&A, 596, A106
Remazeilles, M., Banday, A. J., Baccigalupi, C., et al. 2018, JCAP, 4, 023
Rigby, E. E., Argyle, J., Best, P. N., Rosario, D., & Röttgering, H. J. A. 2015, A&A, 581, A96
Tinti, S., & de Zotti, G. 2006, A&A, 445, 889
Trippe, S., Neri, R., Krips, M., et al. 2010, A&A, 515, A40
Tucci, M., Toffolatti, L., de Zotti, G., & Martínez-González, E. 2011, A&A, 533, A57
Trombetti, T., Burigana, C., De Zotti, G., Galluzzi, V., & Massardi, M. 2018, A&A, 618, A29
Whitehorn, N., Natoli, T., Ade, P. A. R., et al. 2016, ApJ, 830, 143